\def \deg {\ensuremath{^\circ} }

\newcommand*{\fg}[1]{Fig.\thinspace\ref{#1}}
\newcommand*{\fgs}[1]{Figs.\thinspace\ref{#1}}
\newcommand*{\eq}[1]{Eq.\thinspace\ref{#1}}

\documentclass[aps,apl,twocolumn,superscriptaddress]{revtex4-1}

\usepackage{graphicx}% Include figure files
\usepackage{dcolumn}% Align table columns on decimal point

\usepackage{bm}% bold math
\usepackage[english]{babel}

\begin{document}

%___________________________________________________________________________________________
%
% title
%___________________________________________________________________________________________

\title{All-electrical time-resolved spin generation and spin
manipulation in n-InGaAs}% Force line breaks with \\

%___________________________________________________________________________________________
%
% authors
%___________________________________________________________________________________________

\author{I. Stepanov}\affiliation{2nd Institute of Physics and JARA-FIT, RWTH Aachen University, D-52074 Aachen, Germany}
\author{S. Kuhlen}\affiliation{2nd Institute of Physics and JARA-FIT, RWTH Aachen University, D-52074 Aachen, Germany}
\author{M. Ersfeld}\affiliation{2nd Institute of Physics and JARA-FIT, RWTH Aachen University, D-52074 Aachen, Germany}
\author{M. Lepsa}\affiliation{Peter Gr\"{u}nberg Institut (PGI-9) and JARA-FIT, Forschungszentrum J\"{u}lich GmbH, D-52425 J\"{u}lich, Germany}

\author{B. Beschoten}
  \thanks{ e-mail: bernd.beschoten@physik.rwth-aachen.de}
  \affiliation{2nd Institute of Physics and JARA-FIT, RWTH Aachen University, D-52074 Aachen, Germany}

\date{\today}% It is always \today, today, but any date may be explicitly specified

%___________________________________________________________________________________________
%
% abstract
%___________________________________________________________________________________________

\begin{abstract}

 We demonstrate all-electrical spin generation and subsequent manipulation by two successive electric field pulses in an n-InGaAs heterostructure in a time-resolved experiment at zero external magnetic field. The first electric field pulse along the $[1\bar10]$ crystal axis creates a current induced spin polarization (CISP) which is oriented in the plane of the sample. The subsequent electric field pulse along [110] generates a perpendicular magnetic field pulse leading to a coherent precession of this spin polarization with 2-dimensional electrical control over the final spin orientation. Spin precession is probed by time-resolved Faraday rotation. We determine the build-up time of CISP during the first field pulse and extract the spin dephasing time and internal magnetic field strength during the spin manipulation pulse.

\end{abstract}

% PACS, the Physics and Astronomy Classification Scheme.
%Use showkeys class option if keyword display desired
\maketitle

%___________________________________________________________________________________________
% introduction to the field
%

%+ semiconductor and metallic spintronic elements BLA BLA

%+ spin-orbit interaction in semiconductor-heterostructures allows initialization, manipulation and read-out of spin information by electrical means, thus providing basic functional elements of a spintronic device.

%+DC-CISP bla bla. In the same paper pulsed CISP with a photoconductive switch was demonstrated.
%+ electrical manipulation of the spin polarization after optical excitation was first demonstrated in [REF:KATO]. Manipulation with electrical pulses from a pulse-pattern generator and a partial spin rephasing were shown recently in [REF:SEBASTIAN].

%+ the combination of CISP and manipulation with internal magnetic fields in one experiment have been realized in 500nmInGaAs by Kato2008 and Kuhlen2012. Both experiments were conducted with constant voltages and allowed only a limited electrical control over the state of the spin polarization of the system.

%+ pulsed CISP and pulsed Manipulation

Spin-orbit interaction in bulk semiconductors and semiconductor nanostructures provides a variety of useful applications in spintronic devices and can fulfill basic tasks such as electrical initialization, manipulation and detection of electron spin polarizations or spin currents \cite{Review_Awschalom_Spin-based_electronics_vision, Dyankonov, Physics.2.50}.
A charge current in a semiconductor structure can either lead to a homogeneous electron spin polarization (current-induced spin polarization (CISP))
\cite{Edelstein1990233, PRL93_Kato2004_Current-InducedSpinPolarizationinStrainedSemicon%
ductors, NPhys1_Sih2005_SpatialImagingoftheSpinHallEffectandCurrent-InducedPolarizationin2DEG,
PRL97_Stern2006_Current-InducedPolarizationandtheSpinHallEffectatRoomTemperature, APL95_Koehl2009_Current-induced_spin_polarization_in_gallium_nitride}
 or can result in a spin accumulation transverse to the current direction by spin dependent scattering (spin Hall effect) \cite{PhysRevLett.83.1834, PhysRevLett.85.393, Science306_Observation_of_the_Spin_Hall_Effect_in_Semiconductors, NPhys1_Sih2005_SpatialImagingoftheSpinHallEffectandCurrent-InducedPolarizationin2DEG,
PRL94_Wunderlich2005_Experimental_Observation_of_the_Spin-Hall_Effect_in_a_Two-Dimensional,
PRL105_Electrical_Measurement_of_the_Direct_Spin_Hall_Effect_in_Fe-InGaAs_Heterostructures, PhysRevB.88.161305}. Electron spin states can be manipulated by Larmor precession about an effective magnetic field which can be controlled through SO interaction either by static electric fields \cite{Nature427_Kato2004_CoherentSpinManipulationwithoutMagneticField%
sinStrainedSemiconductors,APL87_Electrical_initialization_and_manipulation_of_electron_spins_in_an_L-shaped, NatPhys3_Measurement_of_Rashba_and_Dresselhaus_spin–orbit_magnetic_fields, PRB82_Mapping_spin-orbit_splitting_in_strained_InGaAs_epilayers} or by electric field pulses providing an additional degree of freedom \cite{PRL109_Electric_Field-Driven_Coherent_Spin_Reorientation}. Finally, spin sensitive electrical readout has been demonstrated by the spin-galvanic effect \cite{Nature417_Ganichev2002_Spin-galvaniceffect}. All previous time-resolved experiments on spin initialization and spin manipulation combine time-resolved electrical with ultrafast optical techniques. All-electrical time-resolved experiments on spin initialization and subsequent spin manipulation are still pending.

%___________________________________________________________________________________________
% connection of this paper to the field
In this Letter, we demonstrate that pulsed CISP can be combined with coherent spin manipulation by SO-induced local magnetic field pulses (LMFP) in one time-resolved experiment which allows to achieve all-electrical two-dimensional directional and temporal control of an electron spin polarization in n-InGaAs. The first electric field pulse is applied along the $[1\bar10]$ crystal axis to create an in-plane spin polarization by CISP. The second electric field pulse along the [110] crystal axis triggers a LMFP which leads to coherent Larmor precession of the spin polarization. Time-resolved Faraday rotation (TRFR) is used to probe spin precession during the manipulation pulse (\fg{fig1}(a)). By changing the width and amplitude of both CISP and LMF pulses we achieve all-electrical control over the initial spin polarization, the direction of the Larmor precession and the final spin orientation. Furthermore, the experiment allows a time-resolved detection of the CISP build-up time, the determination of the spin dephasing time and the strength of the LMFP during the spin manipulation pulse.

%___________________________________________________________________________________________
% sample structure
The sample is a 500 nm thick In$_{0.07}$Ga$_{0.93}$As epilayer doped with Si yielding a room temperature carrier density of \mbox{$n\approx 3\times 10^{16}$~cm$^{-3}$} which allows for long spin dephasing times
\cite{PRB66_Dzhioev2002_Low-temperatureSpinRelaxationinN-typeGaAs,PRL80_Kikkawa1998_ResonantSpinAmplificationinN-TypeGaAs, PhysRevLett.105.246603}. The epilayer was grown on a semi-insulating (001) GaAs wafer by molecular beam epitaxy and capped by a 100 nm thick layer of undoped GaAs. The n-InGaAs layer was patterned into a cross-shaped mesa by optical lithography and wet etching (\fg{fig1}(a)). Annealed Au/Ge/Ni electrodes yield ohmic contacts to its center square (200~$\mu$m $\times$ 200~$\mu$m) which provides optical access for spin detection. Each of the four contact pads is connected to a signal line of a coplanar wave-guide. The sample is connected to a dual channel pulse-pattern generator \cite{PRL109_Electric_Field-Driven_Coherent_Spin_Reorientation} and cooled to $T=50$~K in a magneto-optical cryostat \cite{Suppl}.

\begin{figure}[tbp]
\includegraphics{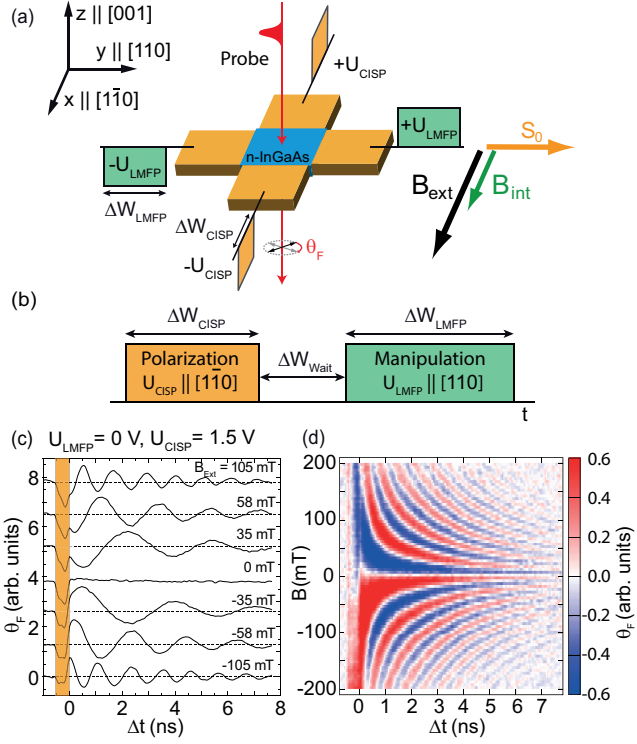}
\caption{ \label{fig1} (color online) (a) Schematic setup. Yellow areas are Au/Ge/Ni contact pads, blue area is the 200~$\mu$m $\times$ 200~$\mu$m InGaAs layer with optical access. The first electric field pulse $\parallel$ $[1\bar10]$ creates a spin polarization which is oriented in the sample plane which is manipulated by a second electric field pulse $\parallel$ $[110]$. Spin precession during the manipulation pulse is probed optically by TRFR. (b) Schematics for pulse sequence of CISP pulse, wait time and LMF spin manipulation pulse with durations of $\Delta W_{\text{CISP}}$, $\Delta W_{\text{wait}}$ and $\Delta W_{\text{LMFP}}$, respectively. (c),(d) TRFR after pulsed CISP excitation in InGaAs at $T=50$~K for various $B_{\text{ext}}$. For all measurements we used CISP pulses with amplitude $U_{\text{CISP}}=+1.5$~V and pulse width $\Delta W_{\text{CISP}}=500$~ps.}
\end{figure}

%___________________________________________________________________________________________
% experiment description
The experiment consists of three basic steps (see \fg{fig1}(a)): (1) generation of an in plane spin polarization along the $y$-direction by an electric field pulse along the $[1\bar10]$ $x$-direction (CISP pulse), (2) coherent spin manipulation by a second electric field pulse along the $[110]$ $y$-direction which launches a LMF pulse with the magnetic field oriented along the $x$-direction and (3) optical detection of the out-of-plane component of the spin polarization along the $z$-direction by measuring time-resolved Faraday rotation $\theta_F$ of a linearly polarized probe pulse. This sample orientation was chosen as it provides strong CISP in $y$-direction (step 1) \cite{PRL93_Kato2004_Current-InducedSpinPolarizationinStrainedSemiconductors} and, at the same time, large electric field-induced internal magnetic fields $B_{\text{int}}$ in $x$-direction \cite{Nature427_Kato2004_CoherentSpinManipulationwithoutMagneticField%
sinStrainedSemiconductors} which are used for spin manipulation (step 2).

As depicted in \fg{fig1}(a), both CISP and LMF pulses each consist of two separate voltage pulse trains of opposite sign ($\pm U_{\text{CISP}}$ for spin polarizing CISP pulse (step 1) and $\pm U_{\text{LMFP}}$ for spin manipulation LMF pulse (step 2)) which arrive simultaneously on opposing sides of the sample. Numerical simulations
\cite{Meier_2007}
show that a geometry with four electrodes can result in rather inhomogeneous electric field distributions. Using two pulses on opposing sides significantly reduces inhomogeneities of the electric field distribution along the InGaAs channel. It furthermore doubles the effective electric field in the center of the sample. The rise-time of the voltage pulses is approximately 100~ps as determined by time-domain reflectometry (not shown). For time-resolved spin detection (step 3), we use an optical pulse train (3~ps pulses) launched from a Ti:sapphire laser. It is synchronized with the pulse-pattern generator at its 80 MHz repetition frequency and has its energy tuned to the fundamental band edge of the InGaAs layer (1.41~eV) \cite{PRL109_Electric_Field-Driven_Coherent_Spin_Reorientation}. Both the current-induced spin polarization pulses and the optical spin detection pulses are modulated for lock-in detection. Spin dephasing times have also independently been determined from all-optical TRFR measurements \cite{PhysRevLett.105.246603}. Thereto, we optically create the initial spin polarization along the $z$-direction by circularly polarized laser pump pulses, which hit the sample under normal incidence ($z$-direction) and use same optical detection scheme as above.

%___________________________________________________________________________________________
% CISP with external field
For time-resolved CISP (step 1) we apply voltage pulses along the x-axis ($[1\bar10]$ direction) with a width of $\Delta W_{\text{CISP}}=500$~ps and an amplitude of of $U_{\text{CISP}}=+1.5$~V which generates an initial in-plane electron spin polarization $S_{\text{0}}$ parallel to the y-axis ($[1\bar10]$ direction, see \fg{fig1}(a)). Furthermore, we apply an external magnetic field $B_{\text{ext}}$ along the x-axis, which is parallel to the electric field but perpendicular to $S_{\text{0}}$. The time-evolution of the out-of-plane component of the resulting spin polarization is measured by TRFR at the center of the sample (\fgs{fig1}(c) and (d)). Most strikingly, we observe several precessions of the spin polarization demonstrating that the voltage pulse triggers a spin polarization with a well-defined phase. Similar results have been obtained by pulsed CISP using a fast photoconductive switch for electrical excitation \cite{PRL93_Kato2004_Current-InducedSpinPolarizationinStrainedSemiconductors}. At $B_{\text{ext}}=0$~T, spins are oriented along the $y$-direction yielding $\theta_F=0$. Spin precession is observed for $B_{\text{ext}}\neq0$~T. Changing the sign of $B_{\text{ext}}$ inverts the direction of spin precession which results in a sign reversal of $\theta_F$. The curves can be described as exponentially decaying oscillations with the Larmor frequency $\omega_L$ proportional to $B_{\text{ext}}$. We note that the initial spin orientation is defined by the CISP pulse while there is no control over the final spin orientation as the spin polarization precesses at all times about the static magnetic field.

%___________________________________________________________________________________________
% CISP and Manipulation
In step 2 we turn off the external magnetic field and manipulate the spin polarization by LMF pulses only. The respective pulse sequence is depicted in \fg{fig1}(b). As above, we generate the initial in-plane spin polarization $S_{\text{0}}\parallel$~$y$ by a CISP pulse ($U_{\text{CISP}}=\pm1.5$~V, $\Delta W_{\text{CISP}}=4$~ns). As there is no external magnetic field, $S_{\text{0}}$ is not precessing during the CISP pulse but instead can reach a larger saturation polarization. A waiting time of $\Delta W_{\text{wait}}=1$~ns after the CISP pulse ensures that a spin-independent background signal from the CISP pulse completely decays before subsequent spin manipulation. The incoming spin manipulation pulse with $U_{\text{LMFP}}=\pm1$~V and $\Delta W_{\text{LMFP}}=7$~ns acts as an internal magnetic field pulse \cite{PRL109_Electric_Field-Driven_Coherent_Spin_Reorientation} with $B_{\text{int}}\parallel$~x (see \fg{fig1}(a)) and rotates $S_{\text{0}}$ out of it's initial in-plane orientation towards the $z$-direction yielding $\theta_F\neq0$. \fg{fig2} shows the time evolutions of $\theta_F$ for various pulse configurations. In \fg{fig2}(a), a positive CISP pulse with $U_{\text{CISP}}=+1.5$V generates a spin polarization which is oriented in the $+y$-direction. The LMF pulse will trigger spin precession in the $zy$ plane during its entire pulse width ($\Delta W_{\text{LMFP}}=7$~ns). Apparently, the spin precession direction reverses when reversing the polarity of the manipulation pulse. This is expected as the direction of the internal magnetic field $B_{\text{int}}$ is reversed at the same time \cite{PRL109_Electric_Field-Driven_Coherent_Spin_Reorientation}. We note that both spin manipulation pulses rotate the spin polarization by $180\deg$ ($\pi$ pulse) from the $+y$ into the $-y$ direction. This gives us a full all-electrical two-dimensional directional control over the spin polarization in the sample.

It is known that the initial spin polarization direction from CISP can be reversed by reversing the CISP voltage \cite{PRL109_Electric_Field-Driven_Coherent_Spin_Reorientation}. For a negative CISP pulse, we therefore generate a spin polarization in the $-y$-direction (\fg{fig2}(b)). Consequently, the respective signs of $\theta_F$ reverses during spin manipulation. No spin precession signals are seen if either the CISP or the LMF pulse are turned off (\fg{fig2}(b)) as in one case no initial spin polarization is generated (green curve) and in the other case the spin polarization stays in-plane (black Curve) and thus cannot be detected by TRFR.

\begin{figure}[tbp]
\includegraphics{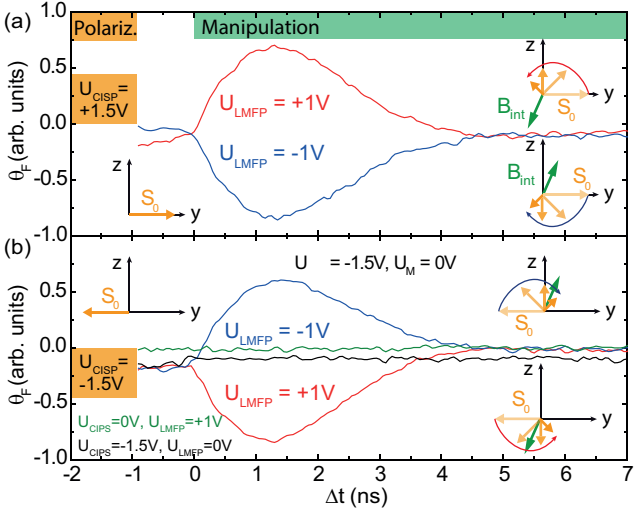}
\caption{ \label{fig2} (color online) TRFR during electrical manipulation for different polarities of $U_{\text{CISP}}=\pm1.5$~V and $U_{\text{LMFP}}=\pm1$~V at $B_{\text{ext}}=0$~mT and $T=50$~K. The TRFR signal changes its sign depending on the initial spin orientation and precession direction which are determined by the polarity of $U_{\text{CISP}}$ and $U_{LMFP}$ (red and blue lines). Evolution of the spin orientation is depicted schematically by yellow arrows. No spin precession is observed if either $U_{\text{CISP}}$ or $U_{\text{LMFP}}$ is set to zero (green and black lines).}
\end{figure}

%___________________________________________________________________________________________
% Variation of pulse lengths
We next focus on the temporal dynamics of the CISP process. The build-up time of CISP has been measured by varying the pulse width $\Delta W_{\text{CISP}}$ of the CISP pulse from 500~ps to 7~ns with the amplitude set to $U_{\text{CISP}}=+1.5$~V. After a waiting time of $\Delta W_{\text{wait}}=1$~ns we apply a subsequent spin manipulation pulse with $U_{\text{LMFP}}=+1.5$~V and $\Delta W_{\text{LMFP}}=4$~ns to probe the temporal evolution of the CISP polarization process. The resulting TRFR curves can be described by an exponentially decaying sine function
\begin{equation}
\theta_F(\Delta t)=\theta_0\cdot\exp\left(-\frac{\Delta t}{T_2^*}\right)\cdot\sin\left(\omega_L \Delta t+\delta\right),
\label{Precession}
\end{equation}
with amplitude $\theta_0\propto S_0$ (CISP), delay $\Delta t$, where we set $\Delta t=0$~ns at the beginning of the spin manipulation pulse at which spin precession starts, spin dephasing time $T_2^*$, Larmor frequency $\omega_L=g
\frac{\mu_B}{\hbar}B_{\text{int}}$ with $\hbar$ being the Planck's constant, $g=$~0.61 the electron g-factor which has independently been determined from time-resolved CISP and all-optical TRFR experiments and the initial phase $\delta=0$ of  the in-plane spin polarization $S_{\text{0}}$.

We fit the data to \eq{Precession} and plot the extracted amplitudes $\theta_0$ against $\Delta W_{\text{CISP}}$ (\fg{fig3}(a), red open squares). The observed exponential increases of $\theta_0$ shows that CISP results from a dynamical process with a build-up time of 1.82 ns. The build-up spin polarization reaches saturation when the build-up rate equals the subsequent spin dephasing rate.

\begin{figure}[tbp]
\includegraphics{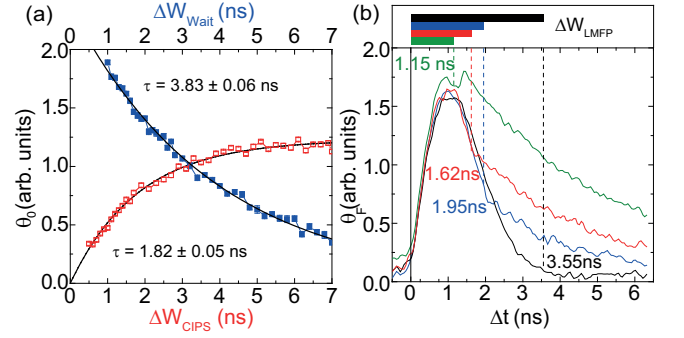}
\caption{ \label{fig3} (color online) (a) Faraday rotation amplitude $\theta_0$ as determined from TRFR during spin manipulation ($U_{\text{LMFP}}$ pulse) using \eq{Precession} vs $\Delta W_{\text{CISP}}$ (red open squares) and $\Delta W_{\text{wait}}$ (blue filled squares), which are respective measures of CISP build-up time during the CISP pulse and the spin relaxation time between the pulses. The time constants $\tau$ are extracted for both processes from an exponential fit to the data. (b) TRFR measurements for different spin manipulation pulse lengths $\Delta W_{\text{LMFP}}$. At shorter $\Delta W_{\text{LMFP}}$ a finite out-of-plane spin polarization remains after the pulse which exponentially decays thereafter.}
\end{figure}

We next determine the in-plane spin lifetime of the CISP by varying $\Delta W_{\text{wait}}$ at fixed $\Delta W_{\text{CISP}}=2$~ns and $\Delta W_{\text{LMFP}}=4$~ns. After the initial spin polarization is generated by the CISP pulse it is given the time $\Delta W_{\text{wait}}$ to decay. The remaining spin polarization is measured afterwards by applying a subsequent LMF pulse which again rotates the spin polarization in the laser probe direction. The dependence of resulting $\theta_0$ on $\Delta W_{\text{wait}}$ is seen in \fg{fig3}(a) (blue filled squares) and shows an exponential decay with a spin lifetime of 3.83~ns. This value is comparable to the out-of-plane spin lifetime of 3.53~ns measured in all-optical pump-probe experiments at $B_{\text{ext}}=0$~mT (see also \fg{fig4}(d)). These results directly confirm the absence of any significant spin relaxation anisotropy between spin orientations along $[110]$ and $[001]$ in this type of heterostructures.

Finally, we vary the width of the spin manipulation pulse $\Delta W_{\text{LMFP}}$ at fixed $\Delta W_{\text{CISP}}=4$~ns and $\Delta W_{\text{wait}}=1$~ns. As the angle of the Larmor precession as well as the total precession time is proportional to the pulse width $\Delta W_{\text{LMFP}}$, we expect to control the final spin orientation by varying $\Delta W_{\text{LMFP}}$. \fg{fig3}(b) shows a sequence of TRFR measurements of the out-of-plane spin polarization for various pulse widths $\Delta W_{\text{LMFP}}$ ranging from 1.15 to 3.55~ns. For the longest pulse width of 3.55~ns the curve is similar to \fg{fig2} and again shows a $\pi$ rotation of the spin polarization. For shorter pulses $\theta_F$ follows the spin precession curve (black curve) during the LMF manipulation pulse. However, spin precession abruptly stops after the manipulation pulse has turned off. This is seen by an exponential decay thereafter. We note that the spin polarization is rotated by $90\deg$ ($\pi/2$ pulse) for $\Delta W_{\text{LMFP}}=1.62$~ns, which is not corresponding to the maximum out-of-plane spin polarization because of relatively short spin lifetime. Along with the control over the initial spin polarization we obtain significant two-dimensional control over the final orientation of the spin polarization.

\begin{figure}[tbp]
\includegraphics{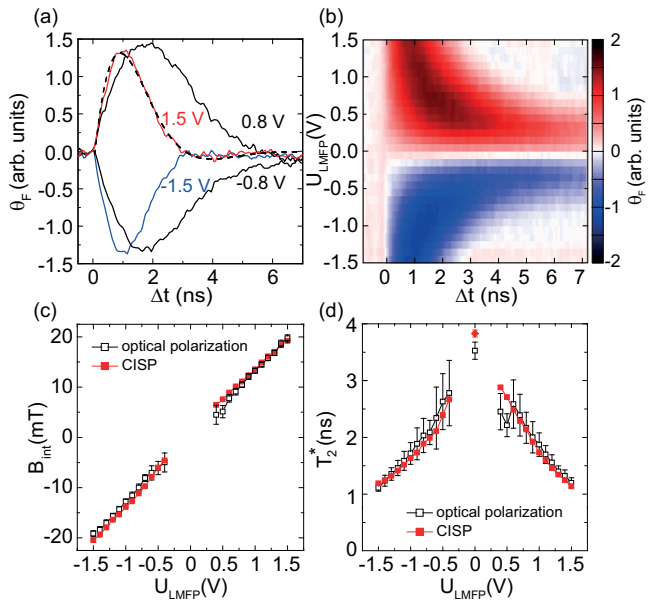}
\caption{ \label{fig4} (color online) (a) and (b) TRFR vs $\Delta t$ for various $U_{\text{LMFP}}$. The dashed line in (a) represent a fit with \eq{Precession} which allows to extract $B_{\text{int}}$ and $T_2^*$. (c) $B_{\text{int}}$ vs $U_{\text{LMFP}}$ after optical (black open squares) and electrical (red filled squares) spin initialization. (d) $T_2^*$ vs $U_{\text{LMFP}}$ after optical (black open squares) and electrical (red filled squares) spin initialization. The red rhombus shows the in-plane-spin relaxation time ($\tau=3.83$~ns) given in \fg{fig3}(a).}
\end{figure}

%___________________________________________________________________________________________
% Variation of manipulation voltage

In all spin manipulation experiments shown in \fgs{fig2} and 3, the LMF pulse only allows for spin rotation by at most $180\deg$ ($\pi$ pulse). This seems to contradict with the time-resolved CISP results in Fig.~1(c) where multiple Larmor precession cycles are visible on much longer time-scales. Apparently, the LMF pulses not only triggers spin precession but also yields significant spin dephasing. We therefore now focus on the magnitude of $B_{\text{int}}$ and the spin dephasing time $T_{2}^{*}$ during coherent spin manipulation.
\fgs{fig4}(a) and (b) show $\theta_F(t)$ during the spin manipulation pulse for different LMF pulse amplitudes $U_{\text{LMFP}}$. The pulse width is set to 7~ns for all measurements. It is obvious that spin precession becomes faster at higher voltages and changes its direction with the polarity of $U_{\text{LMFP}}$. A fit to all measurements by \eq{Precession} allows to extract both voltage dependent spin dephasing times $T_{\text{2}}^{*}$ (red filled squares in Fig.~4(d)) and internal magnetic field strengths $B_{\text{int}}$ (red filled squares in Fig.~4(c))). Similar as in Ref. \cite{PRL109_Electric_Field-Driven_Coherent_Spin_Reorientation}, we observe a linear dependence of $B_{\text{int}}$ on $U_{\text{LMFP}}$ and a voltage induced spin dephasing which is seen by the strong decrease of $T_{2}^{*}$ in Fig.~4(d). The moderate internal magnetic field strength in combination with the short spin dephasing times are the reason that a spin manipulation over $\pi$ by a single pulse is not accessible in present devices. We emphasize, however, that the spin polarization can be rotated into arbitrary directions within the $yz$ plane by using LMF pulses of both positive and negative polarity (see red and blue curve in Fig.~4(a)).

To explore whether the pulsed CISP excitation has any influence on the extracted $B_{\text{int}}$ and $T_{2}^{*}$ values, we have also determined the respective values from a separate set of TRFR measurements where we replace the electrical CISP pulses by circularly polarized laser pulses which are focused onto the sample under normal incidence and thus result in optical spin orientation parallel to the $z$-direction. As in the all-electrical experiment we launch a spin manipulation pulse right after spin excitation. As the internal magnetic field direction is again perpendicular to the spin orientation, it will trigger Larmor precession. Both $B_{\text{int}}$ and $T_{2}^{*}$ are again determined from spin precession during the spin manipulation pulse and are included as black open squares in Figs. 4(c) and (d), respectively.  The results from both measurement techniques are in rather good agreement. The error bar for respective values from optical excitation are larger as our mechanical delay line only covers a shorter time-window of 2.5~ns compared to our long electronic delay of 7~ns which has been used for pulsed CISP excitation.

%___________________________________________________________________________________________
%
% conclusion
%___________________________________________________________________________________________
In conclusion, we have demonstrated that all-electrical electron spin generation and manipulation by two subsequent electrical pulses is achievable in n-InGaAs. High degree of all-electrical temporal and directional control over both the initial and final spin orientation is shown. In addition, we measured the build-up time of CISP for the first time proving that CISP is a dynamical process. Furthermore, we extract internal magnetic fields and spin dephasing times which are identical to all-optical reference experiments demonstrating that pulsed electrical spin polarization, i.e. time-resolved CISP, does not affect these values. The excellent quantitative agreement between both techniques proves the applicability and flexibility of pulsed CISP for spintronic applications.

We gratefully acknowledge financial support from the DFG via FOR 912 (project BE 2441/6-2).


\begin{thebibliography}{25}%
\makeatletter
\providecommand \@ifxundefined [1]{%
 \@ifx{#1\undefined}
}%
\providecommand \@ifnum [1]{%
 \ifnum #1\expandafter \@firstoftwo
 \else \expandafter \@secondoftwo
 \fi
}%
\providecommand \@ifx [1]{%
 \ifx #1\expandafter \@firstoftwo
 \else \expandafter \@secondoftwo
 \fi
}%
\providecommand \natexlab [1]{#1}%
\providecommand \enquote  [1]{``#1''}%
\providecommand \bibnamefont  [1]{#1}%
\providecommand \bibfnamefont [1]{#1}%
\providecommand \citenamefont [1]{#1}%
\providecommand \href@noop [0]{\@secondoftwo}%
\providecommand \href [0]{\begingroup \@sanitize@url \@href}%
\providecommand \@href[1]{\@@startlink{#1}\@@href}%
\providecommand \@@href[1]{\endgroup#1\@@endlink}%
\providecommand \@sanitize@url [0]{\catcode `\\12\catcode `\$12\catcode
  `\&12\catcode `\#12\catcode `\^12\catcode `\_12\catcode `\%12\relax}%
\providecommand \@@startlink[1]{}%
\providecommand \@@endlink[0]{}%
\providecommand \url  [0]{\begingroup\@sanitize@url \@url }%
\providecommand \@url [1]{\endgroup\@href {#1}{\urlprefix }}%
\providecommand \urlprefix  [0]{URL }%
\providecommand \Eprint [0]{\href }%
\providecommand \doibase [0]{http://dx.doi.org/}%
\providecommand \selectlanguage [0]{\@gobble}%
\providecommand \bibinfo  [0]{\@secondoftwo}%
\providecommand \bibfield  [0]{\@secondoftwo}%
\providecommand \translation [1]{[#1]}%
\providecommand \BibitemOpen [0]{}%
\providecommand \bibitemStop [0]{}%
\providecommand \bibitemNoStop [0]{.\EOS\space}%
\providecommand \EOS [0]{\spacefactor3000\relax}%
\providecommand \BibitemShut  [1]{\csname bibitem#1\endcsname}%
\let\auto@bib@innerbib\@empty
%</preamble>
\bibitem [{\citenamefont {Wolf}\ \emph {et~al.}(2001)\citenamefont {Wolf},
  \citenamefont {Awschalom}, \citenamefont {Buhrman}, \citenamefont {Daughton},
  \citenamefont {von Molnár}, \citenamefont {Roukes}, \citenamefont
  {Chtchelkanova},\ and\ \citenamefont
  {Treger}}]{Review_Awschalom_Spin-based_electronics_vision}%
  \BibitemOpen
  \bibfield  {author} {\bibinfo {author} {\bibfnamefont {S.~A.}\ \bibnamefont
  {Wolf}}, \bibinfo {author} {\bibfnamefont {D.~D.}\ \bibnamefont {Awschalom}},
  \bibinfo {author} {\bibfnamefont {R.~A.}\ \bibnamefont {Buhrman}}, \bibinfo
  {author} {\bibfnamefont {J.~M.}\ \bibnamefont {Daughton}}, \bibinfo {author}
  {\bibfnamefont {S.}~\bibnamefont {von Molnár}}, \bibinfo {author}
  {\bibfnamefont {M.~L.}\ \bibnamefont {Roukes}}, \bibinfo {author}
  {\bibfnamefont {A.~Y.}\ \bibnamefont {Chtchelkanova}}, \ and\ \bibinfo
  {author} {\bibfnamefont {D.~M.}\ \bibnamefont {Treger}},\ }\href {\doibase
  10.1126/science.1065389} {\bibfield  {journal} {\bibinfo  {journal}
  {Science}\ }\textbf {\bibinfo {volume} {294}},\ \bibinfo {pages} {1488}
  (\bibinfo {year} {2001})}\BibitemShut {NoStop}%
\bibitem [{\citenamefont {Dyakonov}(2008)}]{Dyankonov}%
  \BibitemOpen
  \bibfield  {author} {\bibinfo {author} {\bibfnamefont {M.}~\bibnamefont
  {Dyakonov}},\ }\href@noop {} {\emph {\bibinfo {title} {{Spin Physics in
  Semiconductors (Springer Series in Solid-State Sciences, 157)}}}}\ (\bibinfo
  {publisher} {Springer},\ \bibinfo {year} {2008})\BibitemShut {NoStop}%
\bibitem [{\citenamefont {Awschalom}\ and\ \citenamefont
  {Samarth}(2009)}]{Physics.2.50}%
  \BibitemOpen
  \bibfield  {author} {\bibinfo {author} {\bibfnamefont {D.}~\bibnamefont
  {Awschalom}}\ and\ \bibinfo {author} {\bibfnamefont {N.}~\bibnamefont
  {Samarth}},\ }\href {\doibase 10.1103/Physics.2.50} {\bibfield  {journal}
  {\bibinfo  {journal} {Physics}\ }\textbf {\bibinfo {volume} {2}},\ \bibinfo
  {pages} {50} (\bibinfo {year} {2009})}\BibitemShut {NoStop}%
\bibitem [{\citenamefont {Edelstein}(1990)}]{Edelstein1990233}%
  \BibitemOpen
  \bibfield  {author} {\bibinfo {author} {\bibfnamefont {V.}~\bibnamefont
  {Edelstein}},\ }\href {\doibase
  http://dx.doi.org/10.1016/0038-1098(90)90963-C} {\bibfield  {journal}
  {\bibinfo  {journal} {Solid State Communications}\ }\textbf {\bibinfo
  {volume} {73}},\ \bibinfo {pages} {233 } (\bibinfo {year}
  {1990})}\BibitemShut {NoStop}%
\bibitem [{\citenamefont {Kato}\ \emph
  {et~al.}(2004{\natexlab{a}})\citenamefont {Kato}, \citenamefont {Myers},
  \citenamefont {Gossard},\ and\ \citenamefont
  {Awschalom}}]{PRL93_Kato2004_Current-InducedSpinPolarizationinStrainedSemico%
nductors}%
  \BibitemOpen
  \bibfield  {author} {\bibinfo {author} {\bibfnamefont {Y.~K.}\ \bibnamefont
  {Kato}}, \bibinfo {author} {\bibfnamefont {R.~C.}\ \bibnamefont {Myers}},
  \bibinfo {author} {\bibfnamefont {A.~C.}\ \bibnamefont {Gossard}}, \ and\
  \bibinfo {author} {\bibfnamefont {D.~D.}\ \bibnamefont {Awschalom}},\ }\href
  {\doibase 10.1103/PhysRevLett.93.176601} {\bibfield  {journal} {\bibinfo
  {journal} {Phys. Rev. Lett.}\ }\textbf {\bibinfo {volume} {93}},\ \bibinfo
  {pages} {176601} (\bibinfo {year} {2004}{\natexlab{a}})}\BibitemShut
  {NoStop}%
\bibitem [{\citenamefont {Sih}\ \emph {et~al.}(2005)\citenamefont {Sih},
  \citenamefont {Myers}, \citenamefont {Kato}, \citenamefont {Lau},
  \citenamefont {Gossard},\ and\ \citenamefont
  {Awschalom}}]{NPhys1_Sih2005_SpatialImagingoftheSpinHallEffectandCurrent-Ind%
ucedPolarizationin2DEG}%
  \BibitemOpen
  \bibfield  {author} {\bibinfo {author} {\bibfnamefont {V.}~\bibnamefont
  {Sih}}, \bibinfo {author} {\bibfnamefont {R.~C.}\ \bibnamefont {Myers}},
  \bibinfo {author} {\bibfnamefont {Y.}~\bibnamefont {Kato}}, \bibinfo {author}
  {\bibfnamefont {W.~H.}\ \bibnamefont {Lau}}, \bibinfo {author} {\bibfnamefont
  {A.~C.}\ \bibnamefont {Gossard}}, \ and\ \bibinfo {author} {\bibfnamefont
  {D.~D.}\ \bibnamefont {Awschalom}},\ }\href {\doibase 10.1038/nphys009}
  {\bibfield  {journal} {\bibinfo  {journal} {Nature Phys.}\ }\textbf {\bibinfo
  {volume} {1}},\ \bibinfo {pages} {31} (\bibinfo {year} {2005})}\BibitemShut
  {NoStop}%
\bibitem [{\citenamefont {Stern}\ \emph {et~al.}(2006)\citenamefont {Stern},
  \citenamefont {Ghosh}, \citenamefont {Xiang}, \citenamefont {Zhu},
  \citenamefont {Samarth},\ and\ \citenamefont
  {Awschalom}}]{PRL97_Stern2006_Current-InducedPolarizationandtheSpinHallEffec%
tatRoomTemperature}%
  \BibitemOpen
  \bibfield  {author} {\bibinfo {author} {\bibfnamefont {N.~P.}\ \bibnamefont
  {Stern}}, \bibinfo {author} {\bibfnamefont {S.}~\bibnamefont {Ghosh}},
  \bibinfo {author} {\bibfnamefont {G.}~\bibnamefont {Xiang}}, \bibinfo
  {author} {\bibfnamefont {M.}~\bibnamefont {Zhu}}, \bibinfo {author}
  {\bibfnamefont {N.}~\bibnamefont {Samarth}}, \ and\ \bibinfo {author}
  {\bibfnamefont {D.~D.}\ \bibnamefont {Awschalom}},\ }\href {\doibase
  10.1103/PhysRevLett.97.126603} {\bibfield  {journal} {\bibinfo  {journal}
  {Phys. Rev. Lett.}\ }\textbf {\bibinfo {volume} {97}},\ \bibinfo {eid}
  {126603} (\bibinfo {year} {2006})}\BibitemShut {NoStop}%
\bibitem [{\citenamefont {Koehl}\ \emph {et~al.}(2009)\citenamefont {Koehl},
  \citenamefont {Wong}, \citenamefont {Poblenz}, \citenamefont {Swenson},
  \citenamefont {Mishra}, \citenamefont {Speck},\ and\ \citenamefont
  {Awschalom}}]{APL95_Koehl2009_Current-induced_spin_polarization_in_gallium_n%
itride}%
  \BibitemOpen
  \bibfield  {author} {\bibinfo {author} {\bibfnamefont {W.~F.}\ \bibnamefont
  {Koehl}}, \bibinfo {author} {\bibfnamefont {M.~H.}\ \bibnamefont {Wong}},
  \bibinfo {author} {\bibfnamefont {C.}~\bibnamefont {Poblenz}}, \bibinfo
  {author} {\bibfnamefont {B.}~\bibnamefont {Swenson}}, \bibinfo {author}
  {\bibfnamefont {U.~K.}\ \bibnamefont {Mishra}}, \bibinfo {author}
  {\bibfnamefont {J.~S.}\ \bibnamefont {Speck}}, \ and\ \bibinfo {author}
  {\bibfnamefont {D.~D.}\ \bibnamefont {Awschalom}},\ }\href {\doibase
  10.1063/1.3194781} {\bibfield  {journal} {\bibinfo  {journal} {Appl. Phys.
  Lett.}\ }\textbf {\bibinfo {volume} {95}},\ \bibinfo {eid} {072110} (\bibinfo
  {year} {2009})}\BibitemShut {NoStop}%
\bibitem [{\citenamefont {Hirsch}(1999)}]{PhysRevLett.83.1834}%
  \BibitemOpen
  \bibfield  {author} {\bibinfo {author} {\bibfnamefont {J.~E.}\ \bibnamefont
  {Hirsch}},\ }\href {\doibase 10.1103/PhysRevLett.83.1834} {\bibfield
  {journal} {\bibinfo  {journal} {Phys. Rev. Lett.}\ }\textbf {\bibinfo
  {volume} {83}},\ \bibinfo {pages} {1834} (\bibinfo {year}
  {1999})}\BibitemShut {NoStop}%
\bibitem [{\citenamefont {Zhang}(2000)}]{PhysRevLett.85.393}%
  \BibitemOpen
  \bibfield  {author} {\bibinfo {author} {\bibfnamefont {S.}~\bibnamefont
  {Zhang}},\ }\href {\doibase 10.1103/PhysRevLett.85.393} {\bibfield  {journal}
  {\bibinfo  {journal} {Phys. Rev. Lett.}\ }\textbf {\bibinfo {volume} {85}},\
  \bibinfo {pages} {393} (\bibinfo {year} {2000})}\BibitemShut {NoStop}%
\bibitem [{\citenamefont {Kato}\ \emph
  {et~al.}(2004{\natexlab{b}})\citenamefont {Kato}, \citenamefont {Myers},
  \citenamefont {Gossard},\ and\ \citenamefont
  {Awschalom}}]{Science306_Observation_of_the_Spin_Hall_Effect_in_Semiconducto%
rs}%
  \BibitemOpen
  \bibfield  {author} {\bibinfo {author} {\bibfnamefont {Y.~K.}\ \bibnamefont
  {Kato}}, \bibinfo {author} {\bibfnamefont {R.~C.}\ \bibnamefont {Myers}},
  \bibinfo {author} {\bibfnamefont {A.~C.}\ \bibnamefont {Gossard}}, \ and\
  \bibinfo {author} {\bibfnamefont {D.~D.}\ \bibnamefont {Awschalom}},\ }\href
  {\doibase 10.1126/science.1105514} {\bibfield  {journal} {\bibinfo  {journal}
  {Science}\ }\textbf {\bibinfo {volume} {306}},\ \bibinfo {pages} {1910}
  (\bibinfo {year} {2004}{\natexlab{b}})}\BibitemShut {NoStop}%
\bibitem [{\citenamefont {Wunderlich}\ \emph {et~al.}(2005)\citenamefont
  {Wunderlich}, \citenamefont {Kaestner}, \citenamefont {Sinova},\ and\
  \citenamefont
  {Jungwirth}}]{PRL94_Wunderlich2005_Experimental_Observation_of_the_Spin-Hall%
_Effect_in_a_Two-Dimensional}%
  \BibitemOpen
  \bibfield  {author} {\bibinfo {author} {\bibfnamefont {J.}~\bibnamefont
  {Wunderlich}}, \bibinfo {author} {\bibfnamefont {B.}~\bibnamefont
  {Kaestner}}, \bibinfo {author} {\bibfnamefont {J.}~\bibnamefont {Sinova}}, \
  and\ \bibinfo {author} {\bibfnamefont {T.}~\bibnamefont {Jungwirth}},\ }\href
  {\doibase 10.1103/PhysRevLett.94.047204} {\bibfield  {journal} {\bibinfo
  {journal} {Phys. Rev. Lett.}\ }\textbf {\bibinfo {volume} {94}},\ \bibinfo
  {pages} {047204} (\bibinfo {year} {2005})}\BibitemShut {NoStop}%
\bibitem [{\citenamefont {Garlid}\ \emph {et~al.}(2010)\citenamefont {Garlid},
  \citenamefont {Hu}, \citenamefont {Chan}, \citenamefont {Palmstr\o{}m},\ and\
  \citenamefont
  {Crowell}}]{PRL105_Electrical_Measurement_of_the_Direct_Spin_Hall_Effect_in_%
Fe-InGaAs_Heterostructures}%
  \BibitemOpen
  \bibfield  {author} {\bibinfo {author} {\bibfnamefont {E.~S.}\ \bibnamefont
  {Garlid}}, \bibinfo {author} {\bibfnamefont {Q.~O.}\ \bibnamefont {Hu}},
  \bibinfo {author} {\bibfnamefont {M.~K.}\ \bibnamefont {Chan}}, \bibinfo
  {author} {\bibfnamefont {C.~J.}\ \bibnamefont {Palmstr\o{}m}}, \ and\
  \bibinfo {author} {\bibfnamefont {P.~A.}\ \bibnamefont {Crowell}},\ }\href
  {\doibase 10.1103/PhysRevLett.105.156602} {\bibfield  {journal} {\bibinfo
  {journal} {Phys. Rev. Lett.}\ }\textbf {\bibinfo {volume} {105}},\ \bibinfo
  {pages} {156602} (\bibinfo {year} {2010})}\BibitemShut {NoStop}%
\bibitem [{\citenamefont {Hernandez}\ \emph {et~al.}(2013)\citenamefont
  {Hernandez}, \citenamefont {Nunes}, \citenamefont {Gusev},\ and\
  \citenamefont {Bakarov}}]{PhysRevB.88.161305}%
  \BibitemOpen
  \bibfield  {author} {\bibinfo {author} {\bibfnamefont {F.~G.~G.}\
  \bibnamefont {Hernandez}}, \bibinfo {author} {\bibfnamefont {L.~M.}\
  \bibnamefont {Nunes}}, \bibinfo {author} {\bibfnamefont {G.~M.}\ \bibnamefont
  {Gusev}}, \ and\ \bibinfo {author} {\bibfnamefont {A.~K.}\ \bibnamefont
  {Bakarov}},\ }\href {\doibase 10.1103/PhysRevB.88.161305} {\bibfield
  {journal} {\bibinfo  {journal} {Phys. Rev. B}\ }\textbf {\bibinfo {volume}
  {88}},\ \bibinfo {pages} {161305} (\bibinfo {year} {2013})}\BibitemShut
  {NoStop}%
\bibitem [{\citenamefont {Kato}\ \emph
  {et~al.}(2004{\natexlab{c}})\citenamefont {Kato}, \citenamefont {Myers},
  \citenamefont {Gossard},\ and\ \citenamefont
  {Awschalom}}]{Nature427_Kato2004_CoherentSpinManipulationwithoutMagneticFiel%
dsinStrainedSemiconductors}%
  \BibitemOpen
  \bibfield  {author} {\bibinfo {author} {\bibfnamefont {Y.}~\bibnamefont
  {Kato}}, \bibinfo {author} {\bibfnamefont {R.~C.}\ \bibnamefont {Myers}},
  \bibinfo {author} {\bibfnamefont {A.~C.}\ \bibnamefont {Gossard}}, \ and\
  \bibinfo {author} {\bibfnamefont {D.~D.}\ \bibnamefont {Awschalom}},\ }\href
  {\doibase 10.1038/nature02202} {\bibfield  {journal} {\bibinfo  {journal}
  {Nature}\ }\textbf {\bibinfo {volume} {427}},\ \bibinfo {pages} {50}
  (\bibinfo {year} {2004}{\natexlab{c}})}\BibitemShut {NoStop}%
\bibitem [{\citenamefont {Kato}\ \emph {et~al.}(2005)\citenamefont {Kato},
  \citenamefont {Myers}, \citenamefont {Gossard},\ and\ \citenamefont
  {Awschalom}}]{APL87_Electrical_initialization_and_manipulation_of_electron_s%
pins_in_an_L-shaped}%
  \BibitemOpen
  \bibfield  {author} {\bibinfo {author} {\bibfnamefont {Y.~K.}\ \bibnamefont
  {Kato}}, \bibinfo {author} {\bibfnamefont {R.~C.}\ \bibnamefont {Myers}},
  \bibinfo {author} {\bibfnamefont {A.~C.}\ \bibnamefont {Gossard}}, \ and\
  \bibinfo {author} {\bibfnamefont {D.~D.}\ \bibnamefont {Awschalom}},\ }\href
  {\doibase 10.1063/1.1994930} {\bibfield  {journal} {\bibinfo  {journal}
  {Appl. Phys. Lett.}\ }\textbf {\bibinfo {volume} {87}},\ \bibinfo {eid}
  {022503} (\bibinfo {year} {2005})}\BibitemShut {NoStop}%
\bibitem [{\citenamefont {Meier}\ \emph {et~al.}(2007)\citenamefont {Meier},
  \citenamefont {Salis}, \citenamefont {Shorubalko}, \citenamefont {Gini},
  \citenamefont {Schön},\ and\ \citenamefont
  {Ensslin}}]{NatPhys3_Measurement_of_Rashba_and_Dresselhaus_spin–orbit_magnet%
ic_fields}%
  \BibitemOpen
  \bibfield  {author} {\bibinfo {author} {\bibfnamefont {L.}~\bibnamefont
  {Meier}}, \bibinfo {author} {\bibfnamefont {G.}~\bibnamefont {Salis}},
  \bibinfo {author} {\bibfnamefont {I.}~\bibnamefont {Shorubalko}}, \bibinfo
  {author} {\bibfnamefont {E.}~\bibnamefont {Gini}}, \bibinfo {author}
  {\bibfnamefont {S.}~\bibnamefont {Schön}}, \ and\ \bibinfo {author}
  {\bibfnamefont {K.}~\bibnamefont {Ensslin}},\ }\href {\doibase
  10.1038/nphys675} {\bibfield  {journal} {\bibinfo  {journal} {Nature
  Physics}\ }\textbf {\bibinfo {volume} {3}},\ \bibinfo {pages} {650} (\bibinfo
  {year} {2007})}\BibitemShut {NoStop}%
\bibitem [{\citenamefont {Norman}\ \emph {et~al.}(2010)\citenamefont {Norman},
  \citenamefont {Trowbridge}, \citenamefont {Stephens}, \citenamefont
  {Gossard}, \citenamefont {Awschalom},\ and\ \citenamefont
  {Sih}}]{PRB82_Mapping_spin-orbit_splitting_in_strained_InGaAs_epilayers}%
  \BibitemOpen
  \bibfield  {author} {\bibinfo {author} {\bibfnamefont {B.~M.}\ \bibnamefont
  {Norman}}, \bibinfo {author} {\bibfnamefont {C.~J.}\ \bibnamefont
  {Trowbridge}}, \bibinfo {author} {\bibfnamefont {J.}~\bibnamefont
  {Stephens}}, \bibinfo {author} {\bibfnamefont {A.~C.}\ \bibnamefont
  {Gossard}}, \bibinfo {author} {\bibfnamefont {D.~D.}\ \bibnamefont
  {Awschalom}}, \ and\ \bibinfo {author} {\bibfnamefont {V.}~\bibnamefont
  {Sih}},\ }\href {\doibase 10.1103/PhysRevB.82.081304} {\bibfield  {journal}
  {\bibinfo  {journal} {Phys. Rev. B}\ }\textbf {\bibinfo {volume} {82}},\
  \bibinfo {pages} {081304} (\bibinfo {year} {2010})}\BibitemShut {NoStop}%
\bibitem [{\citenamefont {Kuhlen}\ \emph {et~al.}(2012)\citenamefont {Kuhlen},
  \citenamefont {Schmalbuch}, \citenamefont {Hagedorn}, \citenamefont
  {Schlammes}, \citenamefont {Patt}, \citenamefont {Lepsa}, \citenamefont
  {G\"untherodt},\ and\ \citenamefont
  {Beschoten}}]{PRL109_Electric_Field-Driven_Coherent_Spin_Reorientation}%
  \BibitemOpen
  \bibfield  {author} {\bibinfo {author} {\bibfnamefont {S.}~\bibnamefont
  {Kuhlen}}, \bibinfo {author} {\bibfnamefont {K.}~\bibnamefont {Schmalbuch}},
  \bibinfo {author} {\bibfnamefont {M.}~\bibnamefont {Hagedorn}}, \bibinfo
  {author} {\bibfnamefont {P.}~\bibnamefont {Schlammes}}, \bibinfo {author}
  {\bibfnamefont {M.}~\bibnamefont {Patt}}, \bibinfo {author} {\bibfnamefont
  {M.}~\bibnamefont {Lepsa}}, \bibinfo {author} {\bibfnamefont
  {G.}~\bibnamefont {G\"untherodt}}, \ and\ \bibinfo {author} {\bibfnamefont
  {B.}~\bibnamefont {Beschoten}},\ }\href {\doibase
  10.1103/PhysRevLett.109.146603} {\bibfield  {journal} {\bibinfo  {journal}
  {Phys. Rev. Lett.}\ }\textbf {\bibinfo {volume} {109}},\ \bibinfo {pages}
  {146603} (\bibinfo {year} {2012})}\BibitemShut {NoStop}%
\bibitem [{\citenamefont {Ganichev}\ \emph {et~al.}(2002)\citenamefont
  {Ganichev}, \citenamefont {Ivchenko}, \citenamefont {Bel'kov}, \citenamefont
  {Tarasenko}, \citenamefont {Sollinger}, \citenamefont {Weiss}, \citenamefont
  {Wegscheider},\ and\ \citenamefont
  {Prettl}}]{Nature417_Ganichev2002_Spin-galvaniceffect}%
  \BibitemOpen
  \bibfield  {author} {\bibinfo {author} {\bibfnamefont {S.~D.}\ \bibnamefont
  {Ganichev}}, \bibinfo {author} {\bibfnamefont {E.~L.}\ \bibnamefont
  {Ivchenko}}, \bibinfo {author} {\bibfnamefont {V.~V.}\ \bibnamefont
  {Bel'kov}}, \bibinfo {author} {\bibfnamefont {S.~A.}\ \bibnamefont
  {Tarasenko}}, \bibinfo {author} {\bibfnamefont {M.}~\bibnamefont
  {Sollinger}}, \bibinfo {author} {\bibfnamefont {D.}~\bibnamefont {Weiss}},
  \bibinfo {author} {\bibfnamefont {W.}~\bibnamefont {Wegscheider}}, \ and\
  \bibinfo {author} {\bibfnamefont {W.}~\bibnamefont {Prettl}},\ }\href
  {\doibase 10.1038/417153a} {\bibfield  {journal} {\bibinfo  {journal}
  {Nature}\ }\textbf {\bibinfo {volume} {417}},\ \bibinfo {pages} {153}
  (\bibinfo {year} {2002})}\BibitemShut {NoStop}%
\bibitem [{\citenamefont {Dzhioev}\ \emph {et~al.}(2002)\citenamefont
  {Dzhioev}, \citenamefont {Kavokin}, \citenamefont {Korenev}, \citenamefont
  {Lazarev}, \citenamefont {Meltser}, \citenamefont {Stepanova}, \citenamefont
  {Zakharchenya}, \citenamefont {Gammon},\ and\ \citenamefont
  {Katzer}}]{PRB66_Dzhioev2002_Low-temperatureSpinRelaxationinN-typeGaAs}%
  \BibitemOpen
  \bibfield  {author} {\bibinfo {author} {\bibfnamefont {R.~I.}\ \bibnamefont
  {Dzhioev}}, \bibinfo {author} {\bibfnamefont {K.~V.}\ \bibnamefont
  {Kavokin}}, \bibinfo {author} {\bibfnamefont {V.~L.}\ \bibnamefont
  {Korenev}}, \bibinfo {author} {\bibfnamefont {M.~V.}\ \bibnamefont
  {Lazarev}}, \bibinfo {author} {\bibfnamefont {B.~Y.}\ \bibnamefont
  {Meltser}}, \bibinfo {author} {\bibfnamefont {M.~N.}\ \bibnamefont
  {Stepanova}}, \bibinfo {author} {\bibfnamefont {B.~P.}\ \bibnamefont
  {Zakharchenya}}, \bibinfo {author} {\bibfnamefont {D.}~\bibnamefont
  {Gammon}}, \ and\ \bibinfo {author} {\bibfnamefont {D.~S.}\ \bibnamefont
  {Katzer}},\ }\href {\doibase 10.1103/PhysRevB.66.245204} {\bibfield
  {journal} {\bibinfo  {journal} {Phys. Rev. B}\ }\textbf {\bibinfo {volume}
  {66}},\ \bibinfo {pages} {245204} (\bibinfo {year} {2002})}\BibitemShut
  {NoStop}%
\bibitem [{\citenamefont {Kikkawa}\ and\ \citenamefont
  {Awschalom}(1998)}]{PRL80_Kikkawa1998_ResonantSpinAmplificationinN-TypeGaAs}%
  \BibitemOpen
  \bibfield  {author} {\bibinfo {author} {\bibfnamefont {J.~M.}\ \bibnamefont
  {Kikkawa}}\ and\ \bibinfo {author} {\bibfnamefont {D.~D.}\ \bibnamefont
  {Awschalom}},\ }\href {\doibase 10.1103/PhysRevLett.80.4313} {\bibfield
  {journal} {\bibinfo  {journal} {Phys. Rev. Lett.}\ }\textbf {\bibinfo
  {volume} {80}},\ \bibinfo {pages} {4313} (\bibinfo {year}
  {1998})}\BibitemShut {NoStop}%
\bibitem [{\citenamefont {Schmalbuch}\ \emph {et~al.}(2010)\citenamefont
  {Schmalbuch}, \citenamefont {G\"obbels}, \citenamefont {Sch\"afers},
  \citenamefont {Rodenb\"ucher}, \citenamefont {Schlammes}, \citenamefont
  {Sch\"apers}, \citenamefont {Lepsa}, \citenamefont {G\"untherodt},\ and\
  \citenamefont {Beschoten}}]{PhysRevLett.105.246603}%
  \BibitemOpen
  \bibfield  {author} {\bibinfo {author} {\bibfnamefont {K.}~\bibnamefont
  {Schmalbuch}}, \bibinfo {author} {\bibfnamefont {S.}~\bibnamefont
  {G\"obbels}}, \bibinfo {author} {\bibfnamefont {P.}~\bibnamefont
  {Sch\"afers}}, \bibinfo {author} {\bibfnamefont {C.}~\bibnamefont
  {Rodenb\"ucher}}, \bibinfo {author} {\bibfnamefont {P.}~\bibnamefont
  {Schlammes}}, \bibinfo {author} {\bibfnamefont {T.}~\bibnamefont
  {Sch\"apers}}, \bibinfo {author} {\bibfnamefont {M.}~\bibnamefont {Lepsa}},
  \bibinfo {author} {\bibfnamefont {G.}~\bibnamefont {G\"untherodt}}, \ and\
  \bibinfo {author} {\bibfnamefont {B.}~\bibnamefont {Beschoten}},\ }\href
  {\doibase 10.1103/PhysRevLett.105.246603} {\bibfield  {journal} {\bibinfo
  {journal} {Phys. Rev. Lett.}\ }\textbf {\bibinfo {volume} {105}},\ \bibinfo
  {pages} {246603} (\bibinfo {year} {2010})}\BibitemShut {NoStop}%
\bibitem{Suppl}
See supplemental material for experimental details.

\bibitem [{\citenamefont {Meier}(2007)}]{Meier_2007}%
  \BibitemOpen
  \bibfield  {author} {\bibinfo {author} {\bibfnamefont {L.}~\bibnamefont
  {Meier}},\ }\emph {\bibinfo {title} {Manipulation of electron spins in
  quantum wells with magnetic and electric fields}},\ \href@noop {} {Ph.D.
  thesis},\ \bibinfo  {school} {ETH Z\"urich} (\bibinfo {year} {2007}),\
  \bibinfo {note} {p. 93}\BibitemShut {NoStop}%
\end{thebibliography}
\end{document}